# Radio-Frequency Gasket for Studies of Superconductivity in Diamond Anvil Cells

Dmitrii V. Semenok[1,*], Di Zhou[1,*], and Viktor V. Struzhkin[2,3,*]

[1] Center for High Pressure Science & Technology Advanced Research, Bldg. #8E, ZPark, 10 Xibeiwang East Rd, Haidian District, Beijing, 100193, China

[2] Shanghai Key Laboratory of Material Frontiers Research in Extreme Environments (MFree), Shanghai Advanced Research in Physical Sciences (SHARPS), 68 Huatuo Rd, Bldg #3 Pudong, Shanghai 201203, China

[3] Center for High Pressure Science & Technology Advanced Research, 1690 Cailun Rd, Bldg #6, Pudong, Shanghai 201203, China

Corresponding authors: Di Zhou (di.zhou@hpstar.ac.cn), Dmitrii V. Semenok (dmitrii.semenok@hpstar.ac.cn), and Viktor V. Struzhkin (viktor.struzhkin@hpstar.ac.cn)

**Abstract**

This work presents the development and testing of a novel radio-frequency (RF) gasket with a Lenz lens surface geometry for contactless measurements in diamond anvil cells (DACs). Conventional RF approaches, which fabricate the Lenz lens onto the diamond anvil itself, preclude the placement of electrical circuits. Our method overcomes this limitation by transferring the RF sensor to a composite Ta-based gasket. The sensor consists of single-turn microcoils that are formed by magnetron sputtering a gold film onto an insulating $Ta_2O_5$ layer. The Lenz lens topology is then patterned using focused ion beam etching. We validated this technique using polycrystalline Cu1234 and Bi2212 high-$T_c$ superconductors at ambient and high pressures. The measurements consistently identified the superconducting transition temperature across carrier frequencies from 111 kHz to 200 MHz. This new gasket technique establishes a reliable and sensitive tool for contactless studies of superconductivity under high pressure.

**Introduction**

    We have previously demonstrated that a Lenz lens can be successfully used not only for NMR studies of superhydrides [1] but also for a simplified method of detecting superconducting, magnetic, and phase transitions [2] by studying the transmission of a radio-frequency (RF) signal through a microscopic sample in high pressure diamond anvil cells (DACs). The implementation of such a lens, however, requires dedicating the surface of a diamond anvil, rendering it unsuitable for other critical roles. We propose a novel solution that relocates this functionality to an "active" gasket. By depositing Lenz lenses on both sides of the gasket, we enable contactless surface impedance studies while freeing the anvils for their primary mechanical and other instrumental functions. Previously, a composite gasket with an insulating layer was used to accommodate electrical contacts for studying the electrical resistance of samples [3, 4]. However, the sensitivity of the van der Pauw scheme to the breakage of even one electrical contact makes this arrangement of electrode placement on the gasket risky. Our approach circumvents this vulnerability by using a system of single-turn coils (a Lenz lens) for inductive studies.

    Using a polycrystalline sample of the cuprate superconductor Cu1234 [5-9], approximately 100 μm in size, placed in a hole in a $Ta_{0.9}W_{0.1}/Ta_2O_5$ gasket, we demonstrate the possibility of directly detecting the superconducting transition temperature ($T_c$), its width, and, possibly, the pseudogap (PG) opening temperature ($T^*$) in this sample. We also describe a method for preparing an insulating gasket with a dense and stable insulating layer based on tantalum ($Ta/Ta_2O_5$) and tantalum-containing alloys for high pressure applications. High pressure tests up to 11 GPa, performed using an optimally doped Bi2212 and Cu1234, allowed us to determine $T_c$ with good accuracy in a sample of about 30 μm in size. The polycrystalline sample of Cu1234 has proven to be a very convenient high-$T_c$ superconducting calibrant for the radio-frequency detection method.

**Results**

*1. Preparation of radio-frequency gasket*

The radio-frequency (RF) gasket was prepared as follows (Figure 1). The initial blank of $Ta_{0.9}W_{0.1}$ alloy (4×4 mm$^2$), containing 10 % of tungsten, was superficially oxidized using an electrochemical anodizing in 0.1M $H_2SO_4$ at a voltage of about 50-60 V for 10 minutes. The surface oxide layer is necessary to reduce the diamond/gasket friction coefficient and prevent the $Ta_{0.9}W_{0.1}$ gasket from seizing during its pre-compression. Besides $Ta_{0.9}W_{0.1}$, we also tested pure Ta gaskets up to pressures of 60-65 GPa, which are quite functional. In this case, the advantage of tantalum is the lack of necessity for electrochemical oxidation of the surface. Finally, a fully insulating gasket can be made from 200-300 μm thick Kapton film with a polymer insert made of B or BN/epoxy as described in Ref. [10]. This type of gasket was used in our experiment at 11 GPa, shown in Figure 8.

After growing the oxide layer, the gasket was pre-indented to a pressure of 20-25 GPa and completely oxidized by heating to 800-1100 ºC in a gas torch flame for 2-5 seconds. The resulting oxide layer ($Ta_2O_5$, Figure 2) exhibits very high adhesion and resistance to mechanical damage. Following this, the edges of the gasket were covered with a protective foil, and the central part of the gasket was coated with 1-2 μm gold layer on both sides using magnetron sputtering. Due to the characteristic roughness of the $Ta_2O_5$ layer, the resulting sputtered layer of Au is highly resistant to mechanical damage. The gold film essentially deposits in the depressions of the composite $Ta_{0.9}W_{0.1}$ or $Ta/Ta_2O_5$ surface and demonstrates high conductivity.

The next preparation step involves laser drilling the gasket in an air atmosphere to oxidize the edges, after the protective foil is removed. A hole substantially larger (1.5-3x) than the diamond culet is drilled to facilitate electrical isolation between the gasket's two surfaces. Although the ideal scenario of complete isolation is not always attained, any residual conductive bridges are typically minimal and have a negligible impact on the acquired RF signal.

The subsequent preparation step involves using Focused Ion Beam (FIB) (Ga or Xe) to create the Lenz lens topology on both surfaces of the gasket. Residual gallium can be removed by etching in an argon beam (3-3.5 keV, 2-5 min). In principle, this could also be achieved by cheaper lithographic methods. It is important to note that the Lenz lens topology implies the presence of one or more radial channels, forming a series of single-turn RF coils. If the gasket is damaged (due to radial cracks), the resulting cracks do not affect the functionality of the RF gasket.

After cutting the conductive layer on both sides of the gasket, two coaxial wires are glued with Ag/epoxy to the inputs of the largest single-turn coil on both sides of the gasket, ensuring electrical isolation and maximum separation of the input and output coaxial channels to reduce mutual inductance and background signal. Finally, the gasket is covered with an additional insulating layer, such as epoxy/$CaF_2$, to exclude electrical contact with heterostructures sputtered onto the diamond anvils. In the test experiment with Cu1234, electrical insulation of the Lenz lenses was not used.

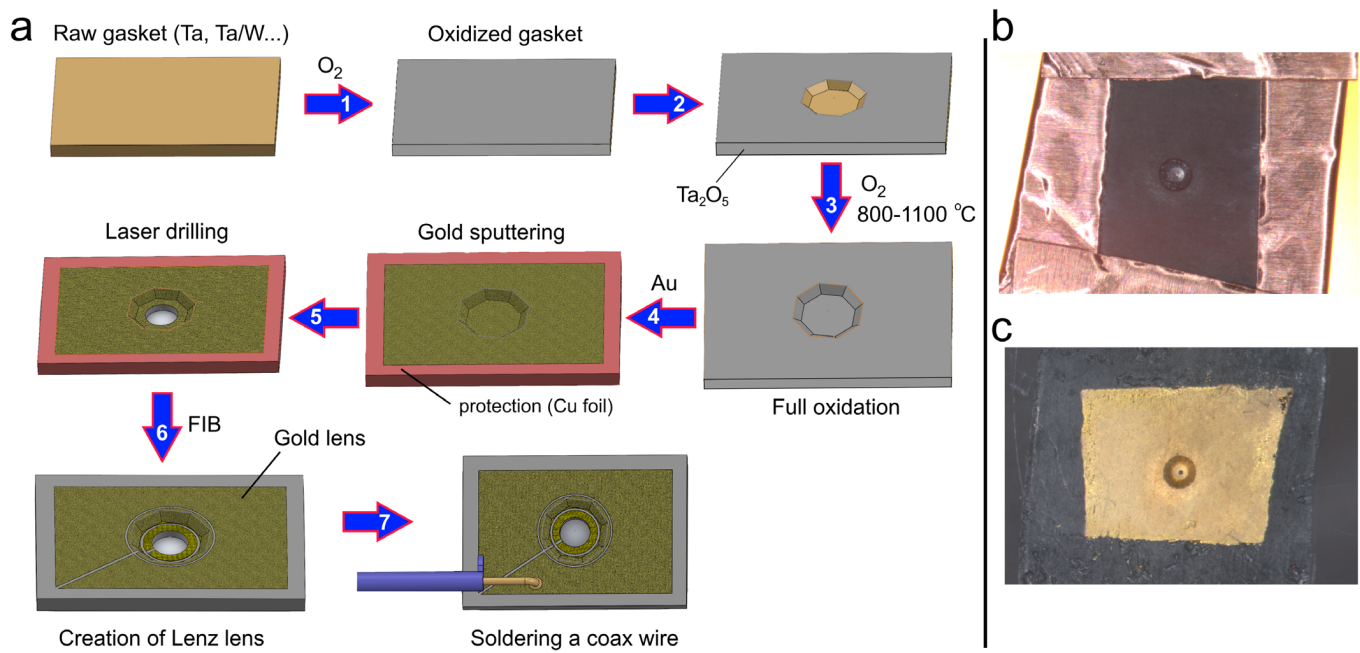

**Figure 1.** Preparation of RF gaskets. (a) Schematic procedure for the preparation of the radio-frequency gasket. The main stages include: 1) Surface oxidation of the Ta-based gasket to prevent seizing during pre-compression; 2) Pre-indentation of the gasket; 3) Full oxidation of the surface to create a durable $Ta_2O_5$ insulating layer, typically by heating in a gas torch flame; 4) Gold sputtering over the $Ta_2O_5$ layer, with the edges protected to maintain isolation; 5) Laser drilling of the central hole in an air atmosphere, which also leads to the oxidation of the hole's edge; 6) FIB etching (or alternative lithographic methods) to create the Lenz lens topology (single-turn microcoils) and removal of the protective layer; and 7) Soldering a coaxial wire to the inputs of the microcoils, ensuring electrical isolation and maximum spatial separation. (b) Photograph of a gasket with copper foil-covered edges. (c) Photograph of the gasket after gold plating and removal of the protective foil.

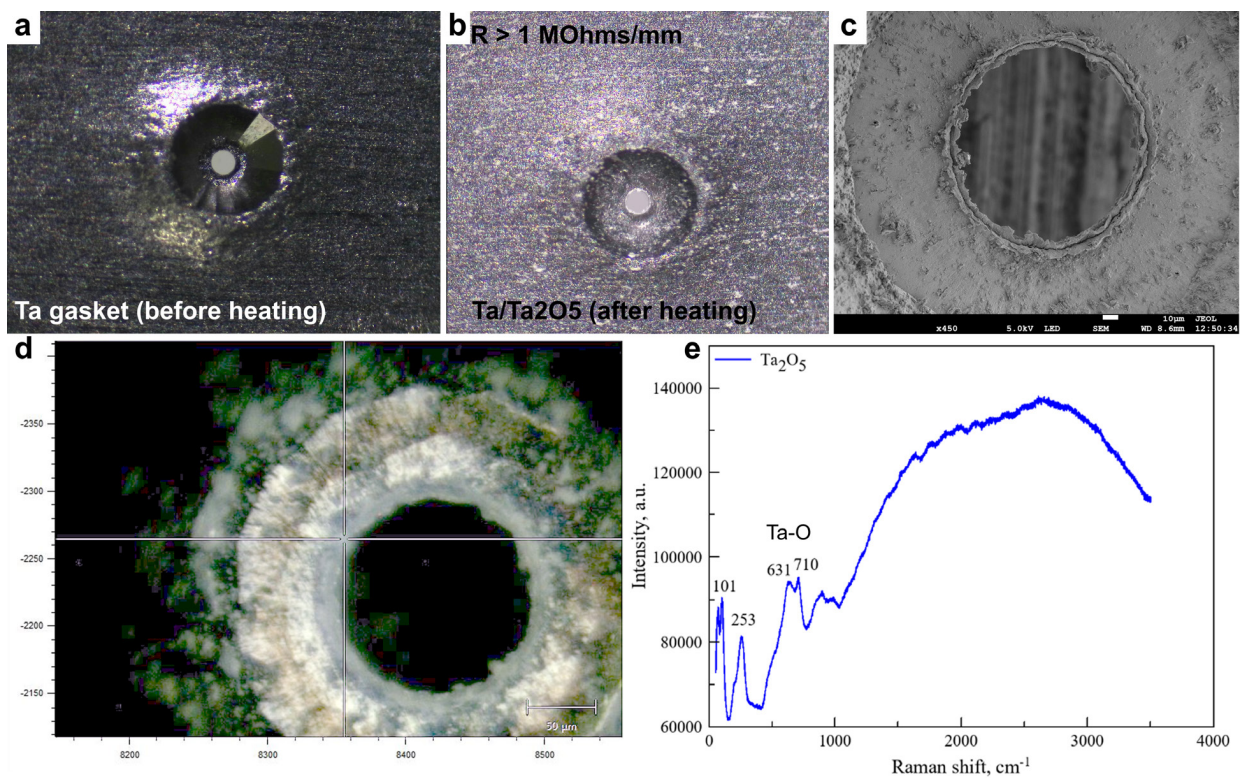

**Figure 2.** Characterization of the tantalum (Ta) gasket with a tantalum pentoxide ($Ta_2O_5$) insulating layer after heating in an air medium. (a) Optical micrograph of the tantalum gasket before heating, showing the sample chamber with a diameter of about 100 μm. (b) Optical micrograph of the tantalum gasket after heating to 800-1100 °C in an air atmosphere, which resulted in the formation of an insulating Ta$_2$O$_5$ layer on the surface. The electrical resistance across the layer is high, R > 1 MΩ for a 1 mm of interelectrode distance. (c) Scanning Electron Microscope (SEM) image of the Ta/Ta$_2$O$_5$ gasket hole, showing the microstructure of the tantalum/oxide layer near the edge. (d) Optical micrograph showing the transparency of the Ta$_2$O$_5$ insulating layer formed at the edge of gasket. (e) Raman spectrum of the insulating material formed on the gasket surface, confirming the presence of Ta$_2$O$_5$. Characteristic vibrational peaks are observed, including those at 101, 253, 631, and 710 cm$^{-1}$, which are consistent with the known vibrational modes of tantalum pentoxide [11].

*2. Study of Cu1234 sample in a test DAC*

Prior to the high-frequency measurements, we performed preliminary electrical resistance *R(T)* characterization of the polycrystalline Cu1234 sample at ambient pressure, with the results shown in Figure 3a. The *R(T)* curve exhibits a very clear superconducting transition, confirmed also by diamagnetic response measurements [5-7], where the resistance drops to near-zero values at low temperatures. We determined an onset critical temperature of about 119 K and an offset critical temperature of 111 K. The finite width of the transition is expected to result in some spread of RF signals from the superconducting transition in the range of 110-120 K, which, as we will see below, is indeed observed.

Furthermore, by examining the first derivative of resistance (*dR/dT*) as a function of temperature (Figure 3b), we identified the manifestation of the PG opening. The normal state resistance deviates from the expected linear trend (indicated by the dashed red line in Figure 3a) starting at a much higher temperature, around $T^* = 148$ K, where the *dR/dT* curve shows a deviation from the linear behavior. This observation confirms the presence of a pseudogap phase preceding the high-$T_c$ superconductivity in the Cu1234 compound, which is common to all cuprates.

Having established the baseline superconducting and PG properties via standard four-probe electrical transport (Figure 3a,b), we proceeded with the radio-frequency transmission experiments to probe the sample's response to the high-frequency electromagnetic fields. For this purpose, a custom RF gasket featuring a fabricated Lenz lens geometry was prepared, as shown in the scanning electron microscope (SEM) image in Figure 3c. The Cu1234 powder was then loaded into the chamber of this RF gasket and placed in the test diamond anvil cell with 100 µm diamond anvil culet (Figure 3d). This radio-frequency gasket replicates the classical geometry proposed by Sakakibara in 1988 for studies of superconductivity in pulsed magnetic fields [12].

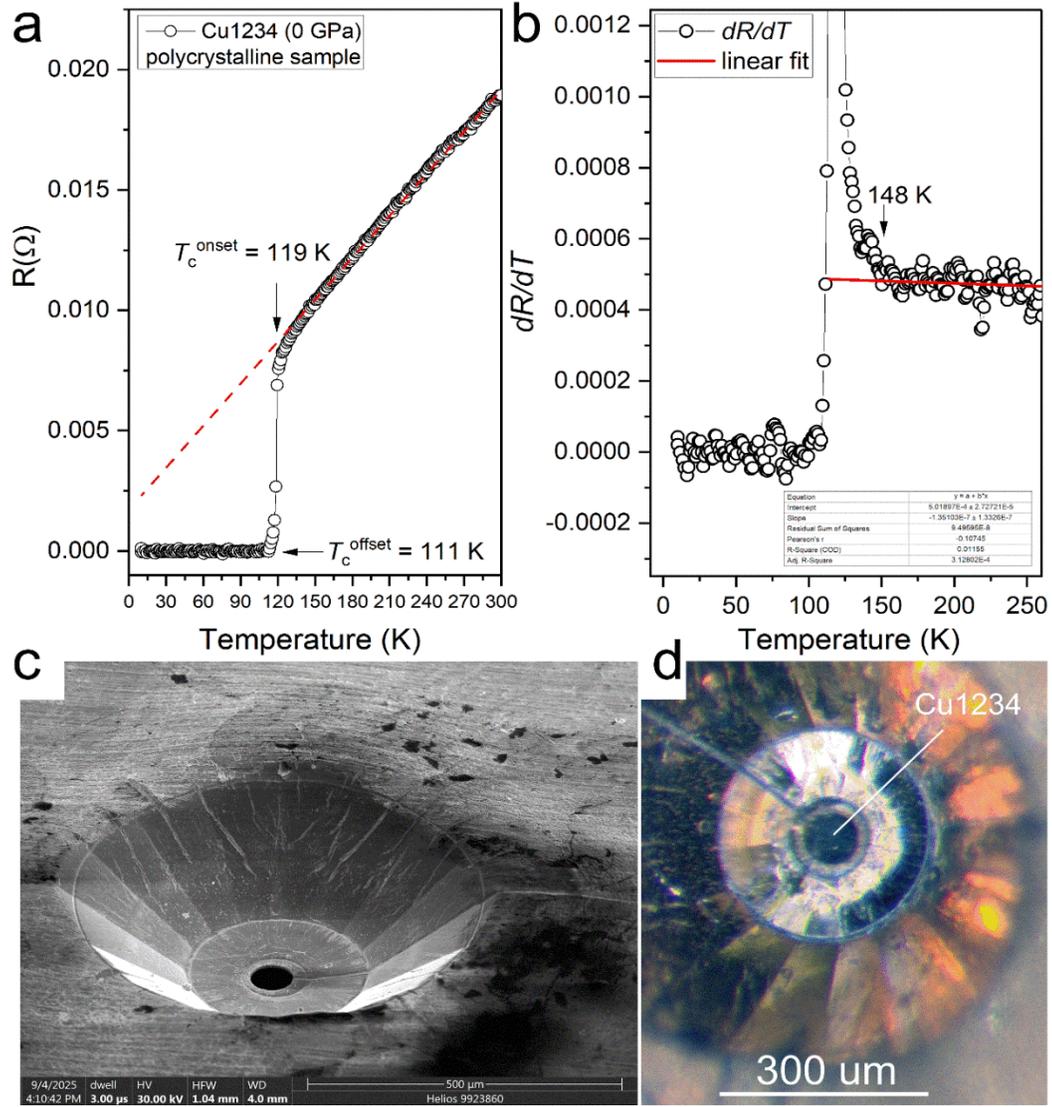

**Figure 3.** Characterization of the Cu1234 superconductor at ambient pressure and loading the RF gasket. (a) Electrical resistance $R(T)$ as a function of temperature for a polycrystalline Cu1234 sample ambient pressure. A clear superconducting transition is observed, with an onset critical temperature ($T_c^{onset}$) of about 119 K, and an offset critical temperature ($T_c^{offset}$) of 111 K. The dashed red line represents a linear fit to the normal state resistance. (b) The first derivative of resistance with respect to temperature ($dR/dT$) as a function of temperature. Deviations of electrical resistance from the linear dependence begin already below 148 K, which indicates the opening of a pseudogap in the compound. A linear fit to the normal state derivative is also shown. (c) Scanning Electron Microscope (SEM) image of the RF gasket with a fabricated Lenz lens geometry used for the experiments. Radial cut near the culet can be seen. (d) Optical micrograph of a Cu1234 sample loaded into the RF gasket, with visible FIB processed paths of the Lenz lens. The scale bar indicates 300 μm.

We conducted radio-frequency transmission measurements on the Cu1234 sample at 0 GPa in a test DAC using a novel experimental setup. The measurement was facilitated by a custom-designed RF gasket described in the previous section (Figures 3c,d and 4a,b). The Cu1234 sample was loaded into the 100 μm chamber (Figures 2c,d) of the RF gasket, enabling the detection of changes in the sample's magnetic permeability and surface conductivity. Figure 4a provides a sketch of the gasket, showing the conductive paths (blue) and the insulating $Ta_2O_5$ separators (brown).

Typical results of the RF transmission measurements at high frequencies of 12.7 and 33.6 MHz, as shown in Figures 4c-e, revealed distinct features as a function of temperature. Specifically, at high carrier frequencies, a surprisingly pronounced signal possibly corresponding to the PG temperature ($T^*$) was observed [13], while the signature of the superconducting transition ($T_c$) was less distinct. For instance, during cooling cycle at carrier frequency of 12.7 MHz (Figure 3c), a clear anomaly indicating the $T_c$ is observed at 115 ± 5 K, but a significantly larger step in both the real and imaginary parts of the signal occurs below $T^* = 137$ K. Certain differences (± 5 K) in the transition temperatures are due to the finite width of the transitions, which is also noticeable in resistive measurements (Figure 3a, b), as well as the hysteresis between warming and cooling cycles.

Measurements taken during warming at a higher frequency of 33.6 MHz (Figures 4d, e) also showed a small drop in the real part corresponding to $T_c = 116 ± 5$ K, overshadowed by a larger peak in both signals below $T^* = 141$-142 K, may possibly be attributed to the pseudogap opening. The clear detection of the superconducting transition suggests that the high-frequency RF technique is highly sensitive to the electronic structure changes associated with this phenomenon in the Cu1234 sample.

To confirm that the measured signal originated solely from the sample and not the measurement setup, we performed a control experiment. We measured the transmitted signal from the same DAC fitted with the same but empty RF gasket, which yielded a featureless signal from 105 K to 150 K (black squares in Figures 4d, e), confirming that the observed at $T_c$ and $T^*$ anomalies are characteristic of the Cu1234 sample.

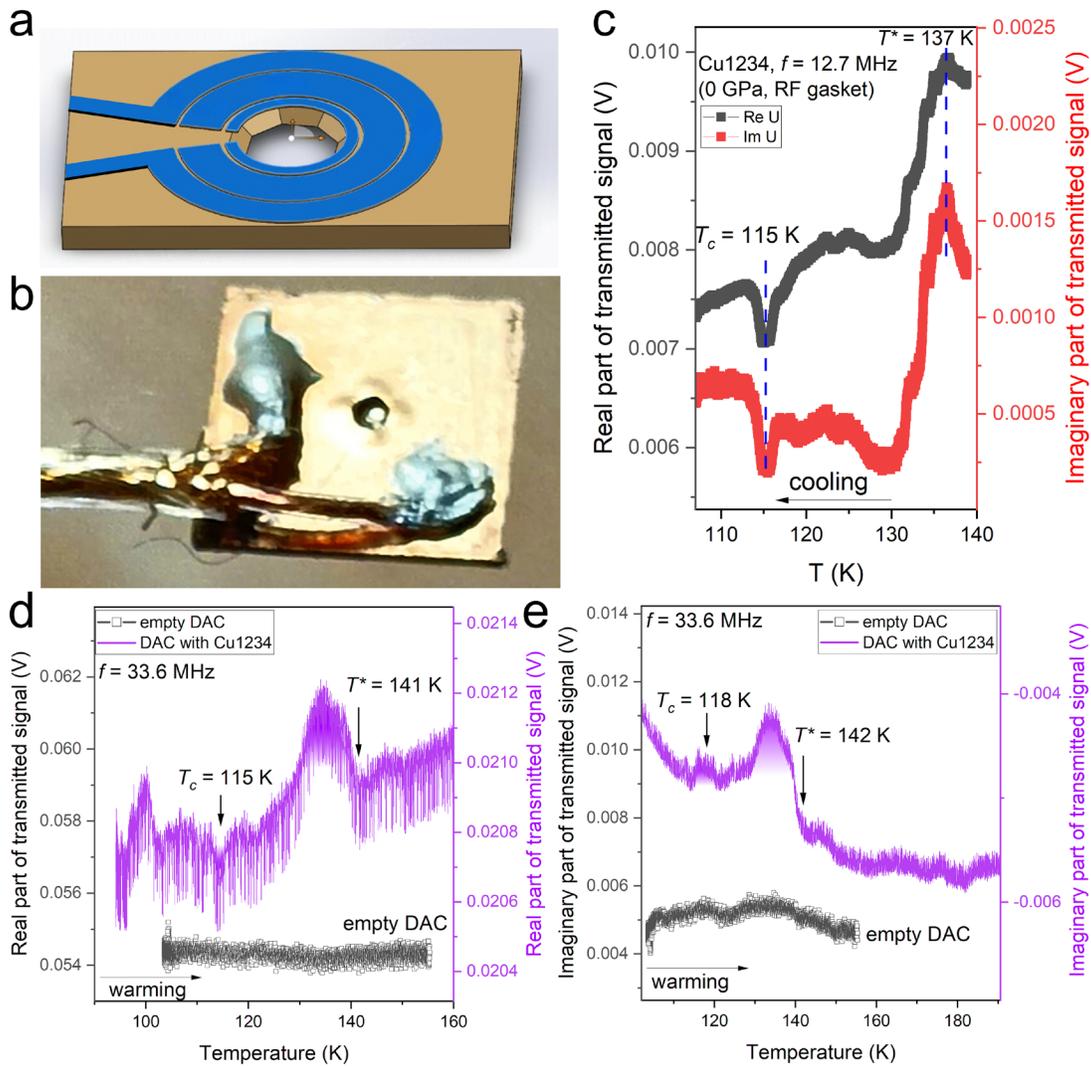

**Figure 4.** Radio-Frequency transmission measurements of a Cu1234 superconductor using a custom RF gasket in a test diamond anvil cell at 0 GPa. (a) A sketch of the custom-designed RF gasket, featuring concentric ring microcoils. The blue parts represent the conductive paths, and the brown represents the insulating oxide separators. (b) Photograph of a real RF gasket with the input coaxial wire connected by silver epoxy glue. (c) Real (black circles) and imaginary (red squares) parts of the transmitted RF signal as a function of temperature during cooling cycle measured at a frequency $f = 12.7$ MHz. A clear anomaly indicating the superconducting transition $T_c$ is observed at 115 ± 5 K, and a larger step in the both parts of the signal corresponding to the PG opening is seen below $T^* = 137$ K. (d, e) Real and imaginary parts of the transmitted RF

signal (purple line) as a function of temperature during warming cycle, measured at the carrier frequency of $f$ = 33.6 MHz. A small drop in the real part signifies the superconducting $T_c$ = 116 ± 5 K, while the larger peak below $T^*$ = 141-142 K may correspond to the opening of the pseudogap. For comparison, the signal from an empty DAC with the same RF gasket is also shown (black squares).

As the carrier signal frequency decreases, the amplitude of the transmitted signal becomes more stable, but at the same time, the relative amplitude of the lock-in detected target signal (for example, a superconducting transition, Figure 5a, b) also decreases as

$$\Delta \mathcal{E} \cong -\Delta \frac{\partial \Phi}{\partial t} \propto -2\pi f N \Delta \left(\frac{\mu}{\rho}\right)_s \times \cos(2\pi f t), \qquad (1)$$

where $\Delta \mathcal{E}$ – is the amplitude of the target signal [2], $\Phi$ is the magnetic flux through the sample, $f$ – is the carrier signal frequency, $N$ - number of turns of the detection coil, $t$ – is the time, and $\Delta \left(\frac{\mu}{\rho_s}\right)$ is a change in the internal properties of the sample: surface resistivity ($\rho_s$), and surface diamagnetic permeability ($\mu_s$). Finally, $\cos(2\pi f t)$ indicates that the target signal is detected by the lock-in amplifier (MFLI, Zurich instruments) at the first harmonic.

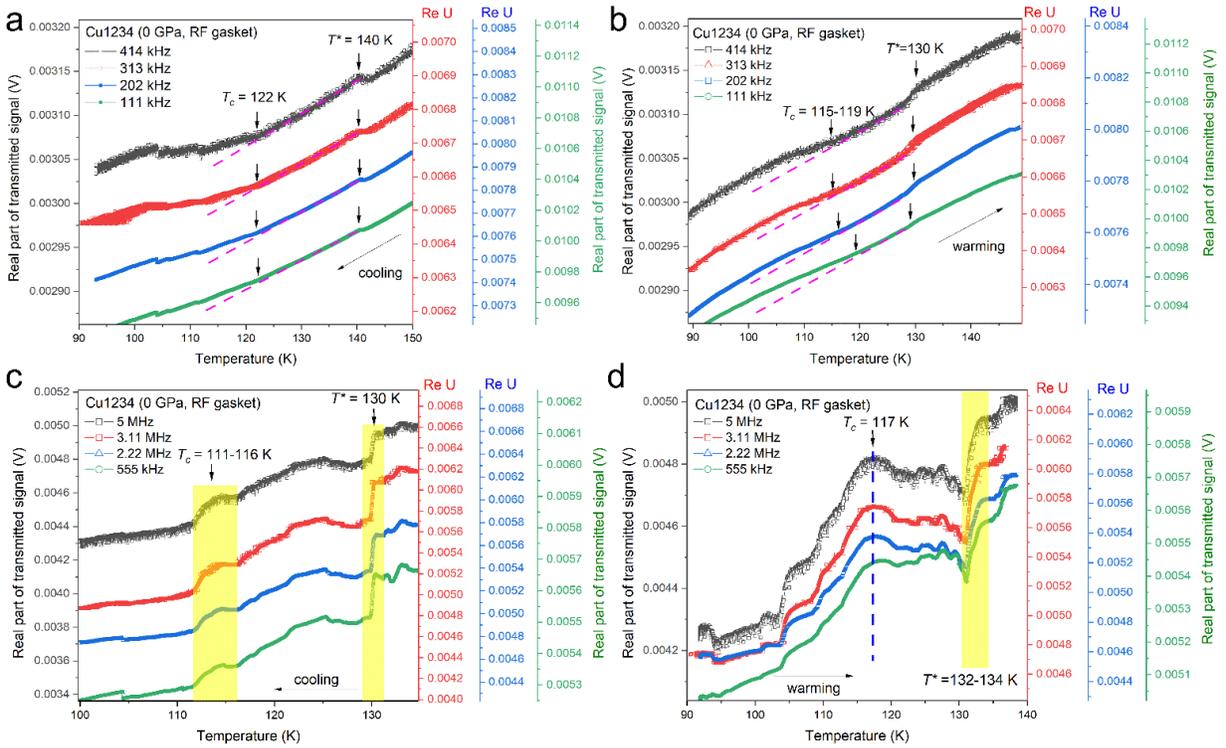

**Figure 5.** Frequency-dependent RF transmission measurements of Cu1234 at 0 GPa using the RF gasket. The panels display the real parts of the transmitted RF signal (Re U) at eight different frequencies, ranging from 5 MHz down to 111 kHz within cooling and warming cycles. (a, b) Real part of the transmitted signal as a function of temperature measured during cooling (a) and warming (b) for a series of lower carrier frequencies (414 kHz, 313 kHz, 202 kHz, 111 kHz). Arrows indicate the superconducting transition ($T_c$) observed at 115–122 K, and an anomaly associated with the pseudogap opening ($T^*$) is evident around 130–140 K. (c, d) The same for higher frequencies 5 MHz, 3.11 MHz, 2.22 MHz, 555 kHz. At high frequencies, transitions become much more pronounced. The superconducting transition occurs between 111–117 K, and the PG temperature is around 130–134 K. The yellow shaded region highlights the superconducting transition and the pseudogap opening region.

As frequencies decrease, the RF transmission method approaches the method of detecting changes in bulk AC diamagnetic permeability. When using this method, reduction of the amplitude of the signal from the sample requires subtracting the background signal from the empty DAC or the signal obtained by extrapolating data above and below the $T_c$ (pink dashed lines). As the frequency increases, the signal becomes more pronounced (Figure 5c, d) and is well reproduced during heating and cooling cycles with an accuracy commensurate with the width of the transitions.

On the other hand, increasing the sample size to several millimeters leads to an increase in its capacitance and inductance, and a decrease in its resonant frequency. In this case, the superconducting transition becomes dominant in the radio-frequency transmission signal, while features in the non-superconducting region are significantly reduced or completely eliminated. In this case, the results of RF measurements are close to those of microwave experiments [14, 15].

*3. High pressure tests of Bi2212 and Cu1234*

The behavior of optimally doped Bi2212 (BSCCO, $Bi_2Sr_2CaCu_2O_{8+\delta}$) under high pressure conditions has been previously studied in several works [16-18]. Typically, as the pressure increases from 0 to ≈10 GPa, the critical temperature of Bi2212 increases from 90-93 K to about 110 K, then begins to decrease, and takes on a near-zero value above 50-55 GPa. In cases of non-stoichiometric samples the situation may change, and the critical temperature decreases almost monotonically with pressure [18]. The pseudogap onset temperature of Bi2212 measured by several methods is $T^* \approx 150$ K at 0 GPa [19, 20]. Considering that the cooling system used in this experiment can only operate down to 78 K, we chose one pressure point (8 GPa) for the high pressure test experiment with an RF gasket, at which $T_c$(Bi2212) is expected to be above 80 K.
A new $Ta/Ta_2O_5$ gasket with a chamber of 50-60 μm in diameter and a sputtered gold layer was used for the experiment, as shown in Figures 1b, c. The diameter of the diamond anvils culet was 30 μm to investigate the feasibility of detecting superconductivity even in a very small sample. After sputtering, a Lenz lens geometry was created on the gaskets using Ga FIB. We initially performed an RF test with an excess of Bi2212, which filled the entire volume of the high pressure chamber after applying of 1-2 GPa (Figure 6a). A study at near-zero pressure in the frequency range from 555 kHz to 5 MHz reveals a well-defined superconducting transition in the form of a step at 92 K (onset, Figure 6b). A kink in the temperature dependence of the transmitted signal is also noticeable at 154 ± 5 K, which may be related to the opening of a pseudogap. At $T' = 100$-$103$ K there is a feature similar to the fluctuation anomaly observed in Ref. [14].

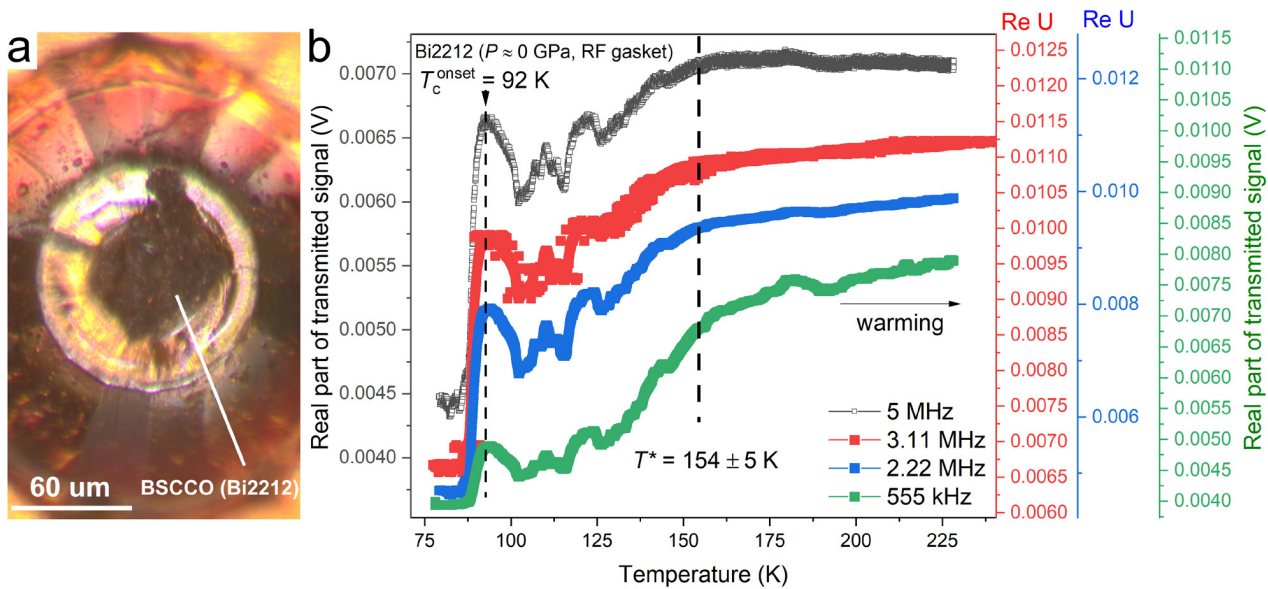

**Figure 6.** The result of the radio-frequency transmission study of a Bi2212 powder microsample of about 60-70 μm in diameter using a gasket with a Lenz lens. (a) Optical photograph of a sample that filled the entire volume of the gasket hole. (b) Real part of the transmitted signal (Re U, in V) as a function of temperature measured during warming for a series of carrier frequencies (555 kHz, 2.22 MHz, 3.11 MHz, 5 MHz).

In the next stage of the experiment, the sample was replaced. A small, thin piece of Bi2212 was exfoliated from the single crystal using adhesive tape and then cut to obtain a particle of approximately 30 μm in diameter (Figure 7a). Paraffin oil was used as the pressure-transmitting medium. At 8 GPa, the particle disintegrated into many small fragments (Figure 7b). A radio-frequency test of the Bi2212 sample using four carrier frequencies ($f_{carr}$ = 5 MHz, 3.11 MHz, 2.22 MHz, and 555 kHz) at 8 GPa demonstrated that the superconductivity signal is observed at ≈ 88 K (Figure 7d,e), which is further confirmed by the presence of a signal at the 2$^{nd}$ harmonic of the modulating magnetic field (Figure 7e).

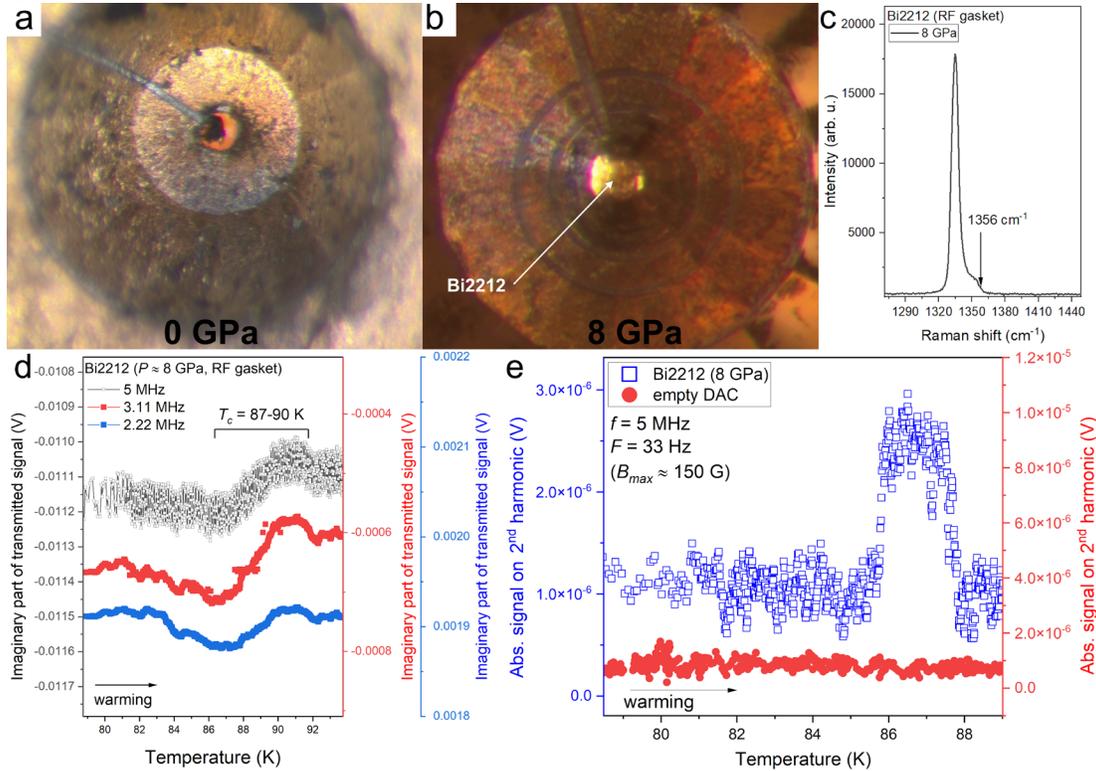

**Figure 7.** Optical microscopy images of a test DAC with RF gasket and loaded Bi2212 sample and radio-frequency tests under pressure using RF gasket. The images correspond to the pressures of (a) 0 GPa, before compression, and (b) 8 GPa. The arrow in (b) points to the Bi2212 sample near the center of the gasket hole. (c) Raman spectra of the DAC's chamber at 8 GPa (black) and 60 GPa (red) in the logarithmic scale. Akahama diamond edge scale was used to determine pressure [21]. (d) Radio-frequency transmission measurements near the superconducting transition at 8 GPa performed during warming cycle at multiple frequencies (5 MHz, 3.11 MHz, 2.22 MHz). The imaginary parts of the transmitted RF signal are shown as a function of temperature. The superconducting transition temperature, $T_c$, is identified around 87-90 K at 8 GPa compared to $T_c$ = 92-94 K at 0 GPa [1] (see also Figure 6). (e) Second harmonic (2$F$ = 66 Hz) signal at 8 GPa: the real part of the signal near $T_c$ for $f_{carr}$ = 5 MHz, and $F$ = 33 Hz. Maximum modulating magnetic field induction was about 150 Gauss. Red points – are signal from the empty DAC with the same RF gasket without any sample.

A similar high pressure test was performed with a Cu1234 sample at 11 GPa in a BX-90 mini non-magnetic cell, equipped with two multi-turn ($N \sim 30$) transmitting and receiving microcoils (Figure 8a). The Lenz lens was made of gold and coated on both sides with a thick Kapton film (200 μm), which served as the outer part of the composite gasket (Figure 8b). A 300 μm diameter hole was laser-drilled in the center of the Kapton gasket and filled with c-BN-filled epoxy resin [10]. After precompression to ~10 GPa and perforation of the high pressure chamber ($d$ = 120 μm), the gasket was covered with gold and patterned using Ga FIB as described in Figure 1. Further testing showed that without using the Lenz lens geometry and magnetic flux focusing, it was impossible to obtain any signal from the superconducting sample placed in the chamber.

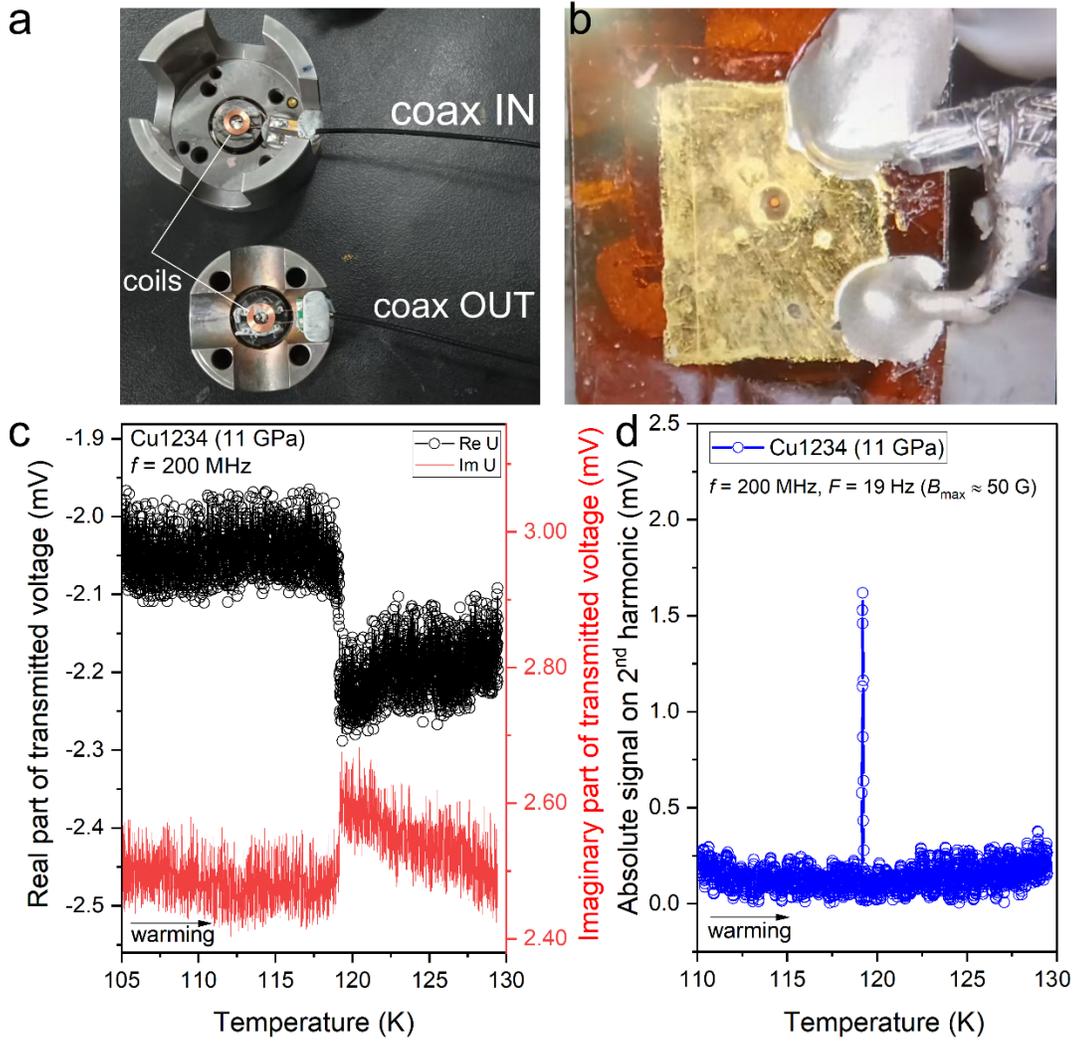

**Figure 8.** Radio-frequency (RF) transmission measurements on Cu1234 cuprate superconductor at 11 GPa using a diamond anvil cell with integrated multiturn coils. (a) Photograph of the RF DAC's piston and cylinder showing two coils connected via coaxial cables labeled "coax IN" and "coax OUT" with a SR844 lock-in amplifier. The coils are wound around diamond anvils. (b) Close-up view of the Kapton gasket with a sputtered gold layer (yellow-gold rectangular part). (c) Temperature-dependent RF transmission data during warming for Cu1234 powder (sample size is about 120 μm) at 11 GPa with carrier frequency $f$ = 200 MHz. The real part of transmitted voltage (Re U, black open circles, left axis) shows a sharp transition near 119-120 K. The imaginary part of transmitted voltage (Im U, red line, right axis) shows corresponding changes with increased noise level. (d) Absolute signal of second harmonic generation at $2F$ (2×19 = 38 Hz) vs. temperature during warming, measured at $f$ = 200 MHz with applied low-frequency magnetic field ($B_{max} \approx$ 50 G). A sharp peak appeared at ~119-120 K corresponds to the superconducting transition temperature.

For the experiment with Cu1234, we used the highest frequency available for the SR844 lock-in amplifier: $f$ = 200 MHz. Although the multi-turn coils at this frequency have significant impedance, limiting the amplitude of the RF signal and the generated electromagnetic field, the useful signal from Cu1234 at 11 GPa is still detectable at the mV level (Figure 8c) due to the multiplier $Nf$ in formula (1). Finally, suppression of the superconducting state by an external AC magnetic field (50 Gauss, $F$ = 19 Hz) near $T = T_c$ leads to oscillations in the high-frequency channel at the second harmonic $2F$ = 38 Hz (Figure 8d). The observed peak on the 2$^{nd}$ harmonic has a width of about $\Delta T$ = 0.5–1 K as in the electrical measurements (Figure 3a), which is also close to $\Delta T$ = 2 K in Bi2212 (Figure 7e), and corresponds to a slightly higher $T_c$ of about 119-120 K, which is consistent with the positive trend of $dT_c/dP > 0$ in Cu1234 found previously [22].

## Conclusions

This study demonstrates a novel "active" gasket that enables robust, contactless inductive radio-frequency measurements in diamond anvil cells. By fabricating Lenz lens microcoils directly onto a composite gasket made of $Ta_{0.9}W_{0.1}$ or $Ta/Ta_2O_5$, as well as Kapton, we transfer the electrical circuits away from the diamond anvils. This innovation not only frees the anvils for auxiliary components like electrodes or thermometers, but also leverages a preparation method that produces a mechanically robust and stable insulating $Ta_2O_5$ layer—a critical feature for high pressure experimentation.

The effectiveness of this RF gasket was shown using polycrystalline Cu1234 and Bi2212 high-$T_c$ cuprate superconductors under ambient and elevated pressure (up to 11 GPa). Our RF transmission measurements of the Cu1234 and Bi2212 microsamples successfully detected the superconducting transition ($T_c$ = 111–122 K), consistent with the classical measurements of the electrical resistance $R(T)$. Furthermore, we confirmed that the signals are genuinely attributable to the Cu1234 and Bi2212 samples by comparing it with a featureless background signal obtained from an empty DAC with RF gasket.

## Acknowledgements

D. V. S. and D. Z. thank the National Natural Science Foundation of China (NSFC, grant No. 12350410354) for the support of this research. D. Z. the Fundamental Research Funds for the Central Universities for support of this research. V.V.S. acknowledges the financial support from Shanghai Science and Technology Committee, China (No. 22JC1410300) and Shanghai Key Laboratory of Materials Frontier Research in Extreme Environments, China (No. 22dz2260800). This work was supported by the National Key Research and Development Program of China (grant 2023YFA1608900, subproject 2023YFA1608903). We thank Prof. Changqing Jin and Dr. Jianfa Zhao (Institute of Physics, CAS) for providing the Cu1234 sample and the electrical resistance data for this experiment.